\documentclass[10pt,a4paper]{article}

\usepackage[utf8]{inputenc}
\usepackage[english]{babel}
\usepackage[T2A]{fontenc}
\usepackage{amsmath}
\usepackage{amsfonts}
\usepackage{amssymb}
\usepackage{graphicx}

\begin{document}

\title{The flow around a macroscopical body by a colloid solution and the drag crisis}

\author{S.V.Iordanski\\
Landau Institute for Theoretical Physics RAS\\
142432 Russia, Chernogolovka}

\maketitle

\begin{abstract}
The motion of colloids in the flow field of a viscous liquid is investigated. The small colloid size
compare to the macroscopical scale of the flow allow to calculate their velocity relative to that of
the liquid. If the inner colloid density is larger then the density of the liquid  the flow field has
the domains where the colloid velocity is close to the liquid velocity. But in the domains with a strong braking of the liquid velocity the colloids are accelerated relative to the liquid. This effect is used for the qualitative explanation of the drag reduction in the flow around macroscopical
bodies and in the pipes.
\end{abstract}

More then 60 years ago \cite{1} it was discovered that a small concentration of 
polymers in liquid solution essentially decrease the drag in pipes. This effect is used in the oil
transportation. There is a lot of theoretical and experimental publications on this subject. However
there is no accepted qualitative interpretation of the physical origin for the observed drug reduction. Rather detailed paper\cite{2} use a complicate theory of the polymer deformation and its
dependence on the inner strain but does not state its connection  with the liquid flow. The recent
work \cite{3} shows a bad agreement of the performed experiment with existing theories. The large
emount of publications is devoted to the rheological properties of the concentrated polymer solutions
see e.g. the review \cite{4}, or the book \cite{5}. We shall not discuss this complicate subject
suggesting that in the dilute polymer solutions the main problem is connected with the interaction
of the separate polymer and the flow field of the liquid. The more subtle effects of the polymer
deformation may be important for more exact quantitative description.

In this note we considier first the more investigated problem of the flow around an immobile 
macroscopical body by Newtonian viscous liquid and its modification due the dilute solution of
comparatively large polymer molecules. The description of the large polymer having thousands of
the connected links is well developed (see e.g.\cite{7} or \cite{8}). The equilibrium state is
represented by a coil having on average a spherical form with the radius $l=\sqrt{\frac{Na^2}{6}}$.
Where $N$ is the number of the links, and $a$ is their length. The molecular weight of such a coil
is much more then the molecular weight of the solvent. Therefore it is possible to neglect the
Brownian motion because the polymer thermal velocity is small to the considering flow velocities.
To simplify the problem further we shall treat the polymer as a large spherical colloid weakly
compressible.
The largest scale of the motion is given by the size of the macroscopic body $L$ which is much more
then the average distance between the nearest  colloids $c^{-1/3}$ where $c$ is the small volume
concentration of the collids. This distance is large compare to the size of the colloid
\begin{equation}
L\gg c^{-1/3}\gg l
\end{equation}

\begin{figure}
 \centering
 \includegraphics[width=5cm]{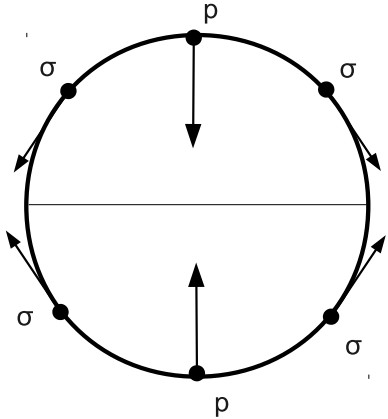}
 \caption{the vertical direction coincides with that of the relative velocity}
\end{figure}

\section{The motion of the colloid relative to the liquid}
At the low colloid concentration it is possible to use the linear approximation (see e.g \cite{9}). Let us consider one colloid in the flow field. At the large distances compare to the colloid size
$l$ the liquid flow can be considered as uniform. The equations of the motion have the form of the
standard equations for an incompressible viscous liquid. We assume that the colloid is incompressible also. At  the colloid boundary the liquid velocities and that of the colloid surface are
equal and the tensor of the momentum transfer is continuous.

The macroscopic motion connected  with the large scale $L$ is a kind of an external force acting on the separate colloid and we can use the well known result (see e.g. \cite{6}) for the calculation
of the force acting on the body (colloid) immersed in the liquid. If the body is moved with the
liquid this force is equal to $\rho^lV_0\frac{dv_i^l}{dt}$, where $\rho^l$ is the density of the liquid, $V_0=\frac{4\pi l^3}{3}$ is the volume of the colloid. But really one must take into account the
relative motion and add $-m_{ik}\frac{d}{dt}(w_i^p-v_i^l)$, where $m_{i,k}$ is the tensor  of the associated masses. The relative motion gives also Stokes friction force 
$-6\pi\eta l(w_i^p-v_i^l)$. Summing the various contributions one get the force acting on the colloid
\begin{equation}
\label{2}
\rho^pV_0\frac{dw_{i}^p}{dt}=\rho^lV_0\frac{dv_i^l}{dt}-m_{ik}\frac{d}{dt}(w_k^p-v_k^l)-
6\pi\eta l(w_i^p-v_i^l)
\end{equation}
 Using Stokes law suggests that the time of the "viscous" relaxation $\frac{l^2\rho^l}{\eta}$ is small compare to the "hydrodynamical" time $L/U$, where $U$ is the velocity of the macroscopic body relative to the liquid. Stokes force is proportional to the first power of the colloid size and therefore the terms with the accelerations are comparatively small. Therefore the zero approximation
gives $w_i^p=v_i^l$ where $v_i^l$ is the local fluid velocity  at a small distance $l$ from the colloid.

The next approximation can be obtained by introducing the zero approximation in \eqref{2} that gives
\begin{equation}
\label{3}
w_i^p-v_i^l=-(\rho^p-\rho^l)\frac{2l^2}{9\eta}\frac{dv_i^l}{dt}
\end{equation}

We shall use this approximation for the relative velocity of the colloid. This local " microscopic"
motion gives the contribution to the averaged equations of motion on the distances of the order $c^{-1/3}$. On
the contrary neglecting this contribution the colloids  have no effect on
the averaged equations of motion.

In Stokes approximation the force acting on unit area of the spherical colloids surface is constant
$F_i=-\frac{3\eta}{l}(w_i^p-v_i^l)$ (see e.g. \cite{6}). Therefore the collid deformations are absent.
In order to find the deformations one needs to consider Oseens corrections connected with the nonlinear inertial
terms. The procedure to find the corrections in Reynolds number 
\begin{equation}
Re=\frac{|\vec{w}^p-\vec{v}^l|}{\eta}\rho^l
\end{equation} 
made in \cite{10,11} can be found in \cite{6}. It is necessary to investigate the solution of the
equation
\begin{equation}
(\vec{u}\vec{\nabla})\vec{v}^l=-\frac{1}{\rho^l}\vec{\nabla}p+\nu\Delta\vec{v}^l
\end{equation}
where $\nu=\frac{\eta}{\rho^l}$ is the kinematic viscosity, $\vec{u}$ is the relative velocity equal
to its constant value far from the coil. In the external flow field there are two domains:
the near one at $r\ll l/Re$ and the far at $r\gg l$, overlapping at $l/{Re}\gg r\gg l$. In the near
domain the starting approximation coincides with Stokes solution, in the far  domain the starting is
Oseen approximation $\vec{u}=const$. The sewing of the appropriate solutions in the overlapping 
region gives
the corrections to Stokes solution $\vec{v}^{(1)}$
\begin{eqnarray}
v_r^{(2)}=\frac{3Re}{8}v_r^{(1)}+\frac{3Re}{32}\left(1-\frac{1}{r'}\right)\left(2+\frac{1}{r'}+\frac{1}{(r')^2}\right)(1-3cos^2\vartheta)\\
v_\vartheta^{(2)}=\frac{3Re}{8}v_{\vartheta}^{(1)}+\frac{3Re}{32}\left(1-\frac{1}{r'}\right)
\left(4+\frac{1}{r'}+\frac{1}{(r')^2}+\frac{2}{(r')^2}\right)sin\vartheta cos\vartheta
\end{eqnarray}
Here the spherical coordinates with the polar ax along the relative velocity are used and the non
dimensional  quantities are introduced:$r'$ in the units of the colloid radius $l$ and the velocities in
the units of the relative velocity $u$. 

The calculations are simplified at small velocities and give at the colloid surface the pressure
\begin{equation}
p^{(2)}=-(1-3cos^2\vartheta)\frac{3}{8l}\eta Re|\vec{u}^l-\vec{w}^p|
\end{equation}
and the tangential strain
\begin{equation}
\label{9}
\sigma_{r\theta}=\frac{3\eta Re}{8l}|\vec{u}^l-\vec{w}^p|sin\vartheta cos\vartheta
\end{equation}

At the picture (1) it is shown some section of the colloid  and the points on its surface with the
maximal stresses and their directions. There is the compression along the relative velocity direction
 and the elongation along the meridians with zeros at the poles and the equator. These stresses
 produce the deformation of the colloid but the calculation require a definite model for the connection
of deformations and stresses inside the colloid. There is the general statement belonging to Maxwell
that the rapid (short time) stresses correspond to the elasticity theory but the slow (long time)
stresses correspond to the viscous liquid. For the case of colloid solutions at high Reynolds numbers
the colloids have the velocity close to the velocity of the solvent that gives the short time
action on the separate colloid in the domain with the large acceleration and most  probable is the validity of the elasticity theory. As a result the colloid is compressed in moving direction and is elongated in
the perpendicular directions. That gives the increased colloid cross section and it moves in the direction
with the largest drag. Therefore the motion instability is possible. However this question has not
direct relation to the drug reduction at the flow around a macroscopic body.

\section{The flow around a macroscopic body.}
The averaged hydrodynamical equations for the incompressible liquid with the solution of  weakly
compressible colloids at the small concentration contain the equation
\begin{equation}
div\vec{v}^l=0
\end{equation}
and the conservation law of the colloid number
\begin{equation}
\frac{\partial c}{\partial t}+div(c\vec{w}^p)=0
\end{equation}

The colloid velocity is close to that of the liquid $\vec{w}^p\approx \vec{v}^l$. Therefore it
follows from the last two equations that $c=const $ is the solution.

The last equation is the conservation law of the total momentum for the system of the liquid and
colloids
\begin{equation}
\label{12}
\frac{\partial}{\partial t}\left[<\rho>v_i^l+cm(w_i^p-v_i^l)\right] =
	-\frac{\partial}{\partial x_k}
	\left[
		\Pi_{ik}+ cm(w_i^p-v_i^l)v_k^l
	\right]
\end{equation}

where $<\rho>=\rho^l+cm$, $\Pi_{ik}=<rho>v_i^l v_k^l +p\delta_{ik}-\eta\frac{\partial v_i^l}{\partial x_k}$, $m=\frac{4\pi l^3}{3}\rho^p$. Stokes drag is absent due to the conservation of the total momentum.

We consider first the absence of the colloids. The general picture of the liquid flow around an immobile
macroscopic body is well investigated. For large Reynolds number $Re=\frac{UL}{\nu}$ at  the far
domain on the distances large compare to the boundary layer thickness the flow is vortex free. At the
boundary layer approximate external surface  the normal to the body velocity is almost zero. The
boundary layer begins in the forward critical point with a finite thickness. Moving in the flow
direction one get the slow  increase of the tangent velocity and the boundary layer thickness till
the maximum of the tangent velocity is achieved. After that the tangent velocity is lowering and
the  laminar  layer become unstable . Its thickness has a sharp increase with a sharp decrease of
the tangent velocity in the same region. This phenomenon is known as "the tearing off" the boundary
layer from the body surface and the creation of the turbulent "stagnation" domain for the flow adjoining to the second critical point as it is shown on fig.2.

\begin{figure}
 \centering
 \includegraphics[width=5cm]{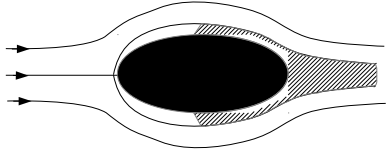}
 \caption{The shaded area is the "stagnation" domain after the "tearing" off }
\end{figure}

If one has the colloid solution in the liquid then the large laminar part of the boundary layer has
a small liquid acceleration and the colloid velocity is quite close to the liquid velocity. Therefore
they do not affect on the average flow according to \eqref{9}. However in the region where the boundary
layer is "tearing off" one has the large acceleration ("braking") at large Reynolds number. That gives the strong colloid stream in the initial flow direction if the colloid density is larger then
that of the liquid according to \eqref{3}.

The drag at large Reynolds numbers in the colloid absence is given (\cite{6}) by formula
$$F=C(Re)1/2 U^2\rho^lS $$ where $S$ is the body cross section, $U$  is the body velocity relative to
the liquid, $C(Re)$ is a constant depending on Reynolds number only. From the experiment it is known
that at the large Reynolds number $Re*$ the function $C(Re)$ essentially changes its form giving
a sharp  drag reduction at comparatively small increase of Reynolds number $\delta Re*\ll Re*$
known as "a drag crisis". This phenomenon is defined as the moving the" tearing off" domain down
the flow and the strong decrease of the "effective" body cross section.

According to \eqref{12} the influence of the colloids in the domain for "tearing off" is quite close to
the enlargement of Reynolds number because the effective momentum stream is increased

\begin{equation}
\Pi_{ik}^{ef}=<\rho>v_i^l v_k^l+p\delta_{ik}-\eta[\frac{\partial v_{i}^l}{\partial x_k}+
\frac{\partial v_k^l}{\partial x_i}+cm\frac{2l^2}{9\eta}\frac{dv_i^l}{dt}v_k^l (\rho^p-\rho^l)
\end{equation}

The colloid contribution can be written as $\delta \Pi_{ik}=\rho^l v_k^l\delta v_i^l$ where
$$\delta v_i^l=cV_0\left(\frac{\rho^p}{\rho^l}\right)^2(1-\frac{\rho^l}{\rho^p})\frac{2l^2}{9\nu}v_n^l\frac{\partial v_i^l}{\partial x_n}$$
The contribution to the integral can be estimated as
$$\int\frac{\delta v_i^l}{\nu}dx\sim cV_0\left(\frac{\rho^p}{\rho^l}\right)^2\left(1-\frac{\rho^l}{\rho^p}\right)\frac{2l^2|v_l|^2}{9\nu^2}$$

This expression gives the estimate of the colloid contribution to Reynolds number. If this quantity
is equal to the experimental $\delta Re*>0$ then one get the drug reduction like the drug crisis
without the polymer solution. The condition of the drug reduction by the polymer solution define 
the coil concentration and the sign of the difference $\rho^p-\rho^l>0$ . This condition can be rewritten in the form

\begin{equation}
\delta Re*=cV_0\left(\frac{\rho^p}{\rho^l}\right)^2\left(1-\frac{\rho^l}{\rho^p}\right)\frac{2l^2}{9L^2}\left(Re*\right)^2
\end{equation}
By suggestion $cV_0\ll 1$ therefore the requirement $\delta Re* \frac{L^2}{(Re*)^2}\ll l^2$ for the
colloid size must be fulfilled. This consideration does not give the difficult numerical estimate.

The details of the turbulent flows in a long pipes at large Reynolds numbers are not so well investigated . Partially phenomenological Karman-Prandtl theory is based (as was shown by L.D.Landau)
on the dimensional analysis and the experimental determination of some empirical constants. The main
statement \cite{6} is the existence of the central area in the pipe cross section with a weak
dependence of the mean velocity  on the radius (fig.3 the shaded area) and the viscous sublayer close to 
the wall of the tube with the linear decrease of the liquid velocity to zero at the wall. The
velocity derivative in the viscous sublayer is large compare to that in the central area.

If the colloids are present the flow in the central domain is weakly affected because the mean
colloid velocities are very close to that of the liquid. However near the transition boundary colloids can achieve an essential velocity according to \eqref{3} if $\rho^p\>\rho^l$. Therefore
the central domain with the large mean velocity must increase due to the colloids motion. The numerical
estimate is difficult because one must know the mean value of the acceleration $\frac{dv_i^l}{dt}$ for
the turbulent flow in the transition region, and the unknown empirical constants defining the central
and side domains.

\begin{figure}
 \centering
 \includegraphics[width=5cm]{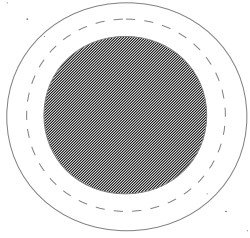}
 \caption{the central domain with a large mean velocity weakly dependent on radius is shaded.
 Its  enlargement is shown by a dotted line. }
\end{figure}

Author express his gratitude to E.Kats, I.Kolokolov and V.Lebedev for the numerous discussions on the
subject. The work is partially supported by the Ministry of Science and Education RF (Russian
Federal Targeted Programs  "S\&SPPIP"and "I\&DPFS\&T".


\begin{thebibliography}{99}
\bibitem{1} B.A.Toms(1949) Proc.Int.Rheological congress,Holland (1948) p.135
\bibitem{2}I.Procaccia,V.Lvov, R.Benzi,arXiv:nlin/0702034v1[nlin.CD]15 Feb.2007 
\bibitem{3}You.Burnishev,V.Steinberg,EPL 100(2012)24001,doi:10.1209/0295-5075/100/24001
\bibitem{4}Suzanne M.Fielding,Softmatter, 2007,3,1262-1279
\bibitem{5}R.G.Larson,Constitutive equations for polimer melts,Butterworth series in chemical
engineering (1988)
\bibitem{6}L.D.Landau,Е.М.Lifshits, HydroDynamics,Moscow ,Physmathlit(2003)
\bibitem{7}M.Kleman,О.Lavrentovich, Soft Matter Physics,Springer 2003 Москва,Физматлит 2007, 
\bibitem{8}М.Doi,S.F.Edwards ,The theory  of polymer dynamics, Мир (1986),Clarendon Press-Oxford
\bibitem{9}С.В.Иорданский,А.Г.Куликовский, Механика жидкости и газа ,(Известия АН СССР) №4,1977,р12
\bibitem{10}S.Kaplun,P.A.Lagerstrom, J.Math. and Phys.6(1957),p.585-593
\bibitem{11}I.Proudman,J.R.A.Pearson, J.Fluid Mech.2(1957)p. 237-262

\end{thebibliography}
\end{document}